\documentstyle[12pt,world_sci,epsf]{article}
\def\abs#1{\left| #1\right|}

\begin{document}
\title{{\bf CDF Results on {\boldmath $B$} Decays}}
\author{JOHN E. SKARHA\thanks{For the CDF Collaboration;
Contribution to the proceedings of the LISHEP95 Workshop on Heavy Flavor
Physics, Rio de Janeiro, Brazil, February 20-22, 1995.}\\
Department of Physics and Astronomy\\
The Johns Hopkins University\\
Baltimore, Maryland 21218, USA}
\vspace{0.3cm}

\maketitle
\setlength{\baselineskip}{2.6ex}

\begin{center}
\parbox{13.0cm}
{\begin{center} ABSTRACT \end{center}
{\small \hspace*{0.3cm} We present recent CDF results on
$B$ lifetimes, $B$ meson mass measurements, ratios of branching ratios,
and rare decays.  In addition, we present the first
measurement of time-dependent $B_d$ mixing at CDF.
Several results have been updated and a few new ones included since
the workshop.}}
\end{center}

\section{Introduction}
During the 1992-95 Tevatron collider Run I, the Collider Detector at
Fermilab (CDF)~\cite{CDF} has so far collected a data sample
of $\overline{p}p$ collisions at $\sqrt{s} = 1.8$ TeV with an integrated
luminosity of $> 95$ pb$^{-1}$.  This was split into two separate
data-taking runs: Run~1a ($\sim 20$ pb$^{-1}$) and the present Run~1b
($> 75$ pb$^{-1}$).  Data-taking is planned to continue through 1995 and
a total sample of at least 120 pb$^{-1}$ is expected.
This
data sample, in combination
with improvements to the data acquisition system, the muon coverage, and
most importantly, the installation of the CDF SVX silicon vertex
detector~\cite{svx},
has allowed many new results on $B$ decays.
In this paper we report results on $B$ lifetimes, ratios of branching ratios,
rare decays, and $B$ meson mass measurements.  In addition, we present
the first CDF measurement of time-dependent $B_d$ mixing.

\section{$B^+$ and $B^0$ Meson Lifetimes}
Increasingly precise measurements of the $B^+$ and $B^0$ lifetimes are
important for testing the predicted $B$ hadron
lifetime hierarchy and to measure the relative
contributions from non-spectator decays.
Only small lifetime differences are expected between the
$B^+$ and $B^0$ mesons ($\sim 5\%$~\cite{thlife}) and experiments are now
approaching this precision.

At CDF, the measurement of the charged and neutral $B$ meson lifetimes has
been performed using fully reconstructed $B$ decays in the following
modes~\cite{charge}:
\begin{displaymath}
\begin{array}[b]{rllrll}
B^+ &\rightarrow J/\psi K^+       &\rightarrow \mu^+\mu^- K^+; &
B^+ &\rightarrow J/\psi K^*(892)^+ &\rightarrow \mu^+\mu^- K^0_S\pi^+ \\
B^+ &\rightarrow \psi (2S) K^+    &\rightarrow \mu^+\mu^-\pi^+\pi^- K^+; &
B^+ &\rightarrow \psi (2S) K^*(892)^+ &\rightarrow \mu^+\mu^-\pi^+\pi^-
K^0_S\pi^+\\
B^0 &\rightarrow J/\psi K^0_S     &\rightarrow \mu^+\mu^- K^0_S; &
B^0 &\rightarrow J/\psi K^*(892)^0    &\rightarrow \mu^+\mu^- K^+\pi^- \\
B^0 &\rightarrow \psi (2S) K^0_S  &\rightarrow \mu^+\mu^-\pi^+\pi^- K^0_S; &
B^0 &\rightarrow \psi (2S) K^*(892)^0 &\rightarrow \mu^+\mu^-\pi^+\pi^-
K^+\pi^-
\end{array}
\end{displaymath}

The capability of using exclusive decay modes is unique to CDF, no other
experiment has large samples of fully reconstructed $B$ decays that can be
used for lifetime measurements.  $B^+$ and $B^0$ lifetime measurements using
exclusive decays have been previously published for the $\sim 20$ pb$^{-1}$
Run~1a data sample~\cite{cdf_exlife}.

CDF has recently added an additional $\sim 48$ pb$^{-1}$ of
$J/\psi \rightarrow \mu^+ \mu^-$ data from Run~1b,
bringing the total data sample to 67.7 pb$^{-1}$.
As in the Run~1a analysis, reconstruction of
$J/\psi \rightarrow \mu^+ \mu^-$ candidates
is the starting point.  The $\psi(2S) \rightarrow J/\psi \pi^+ \pi^-$
decays are
then searched for in that data sample.  Two track combinations are
used to find the $K^*(892)^0$ and $K^0_S$ candidates.
The $\psi(2S)$ and $K^0_S$ candidates are required to be within 20 MeV/c$^2$,
while $J/\psi$ and $K^*$ candidates are
required to be
within 80 MeV/c$^{2}$ of their respective world average values~\cite{pdg}.
The invariant mass distributions of the $J/\psi$ and $\psi(2S)$ are shown
in Figure~1.
\vspace{-0.1in}
\begin{figure}[thb]
\parbox[b]{3.0in}{\epsfxsize=2.9in\epsfbox{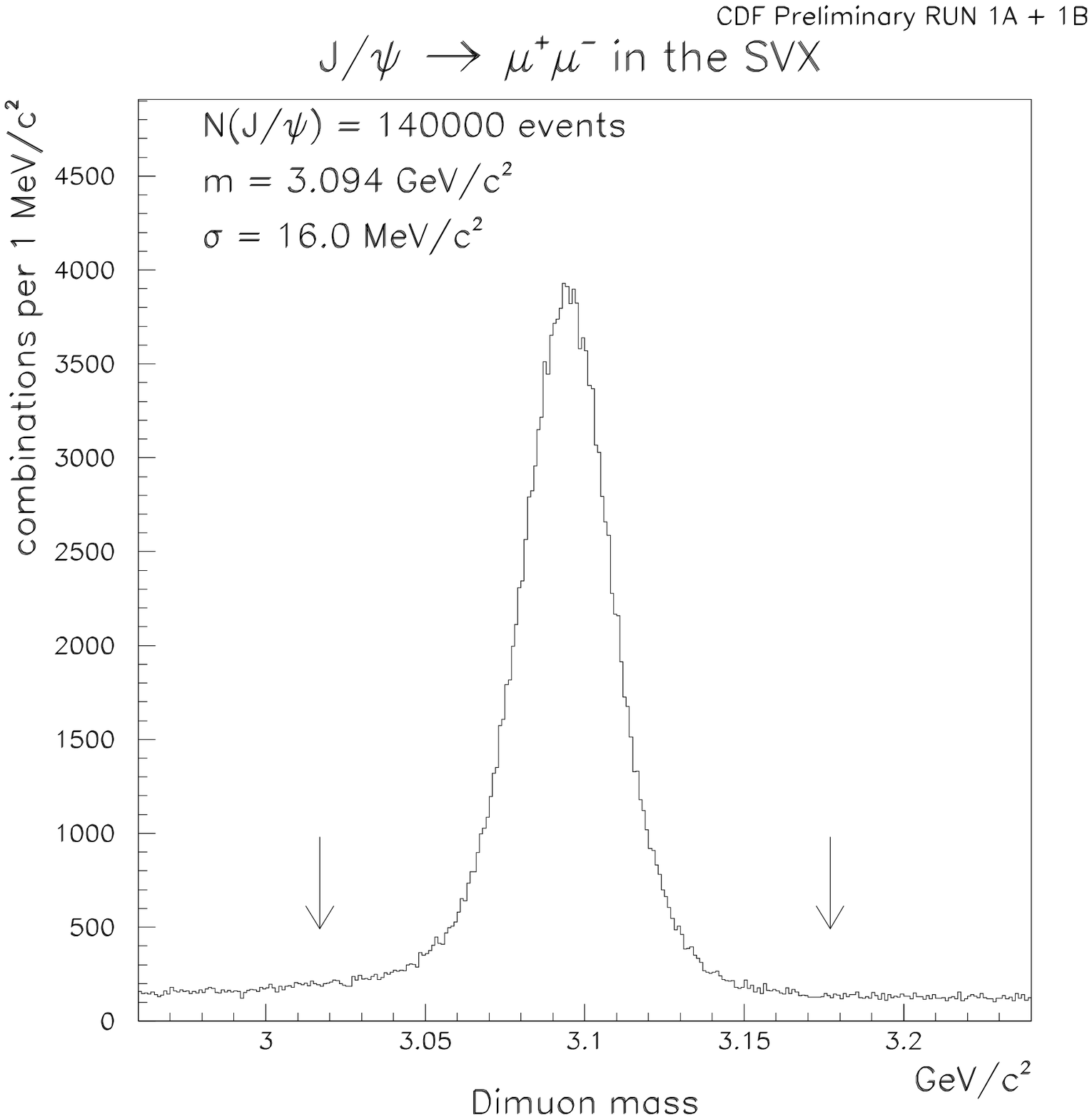}}%
\makebox[0in][r]{\raisebox{3.2in}{\em  a) \hspace{0.75in}}}
\label{psis}
\hfill
\parbox[b]{3.0in}{\epsfxsize=2.9in\epsfbox{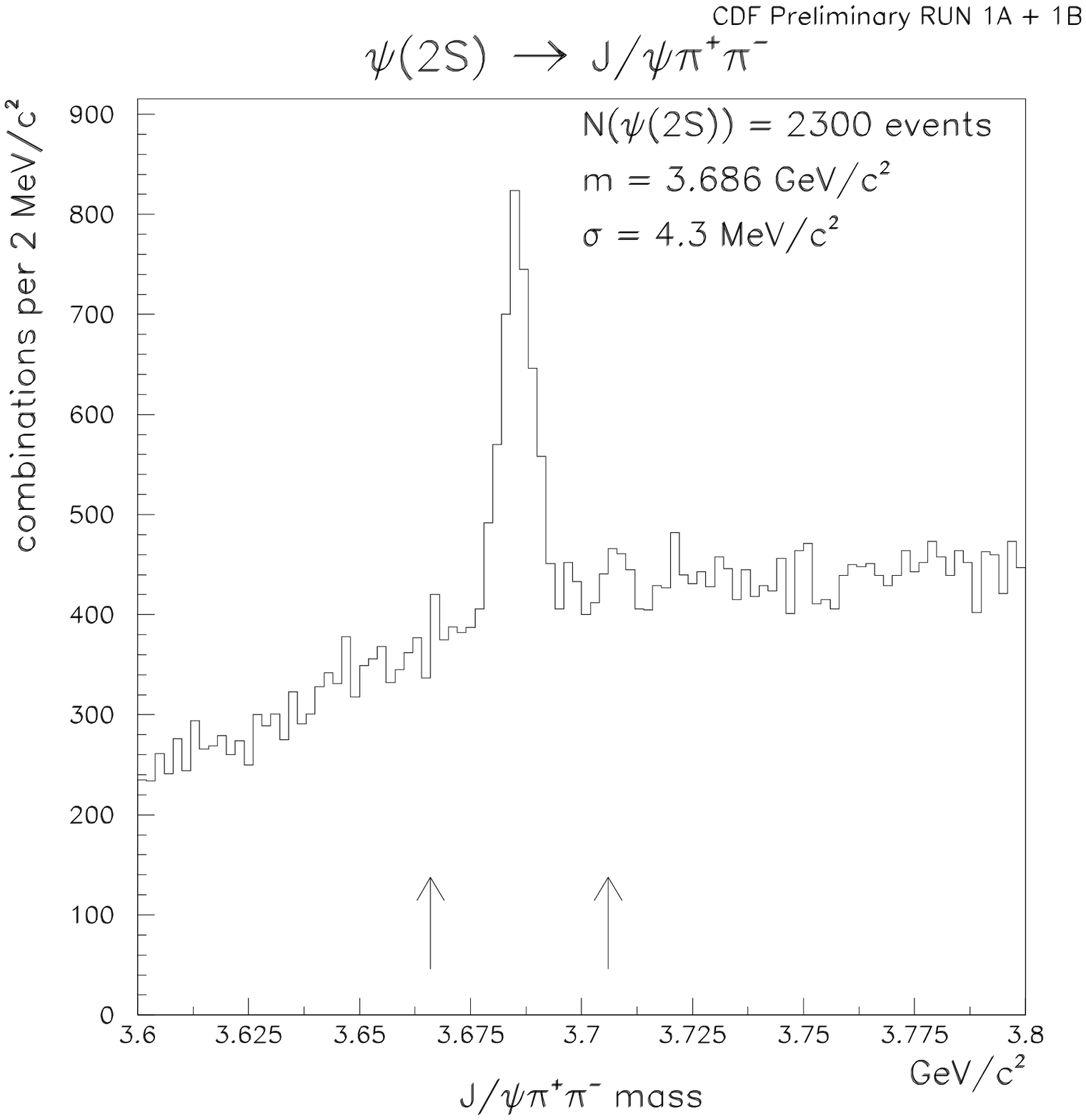}}%
\makebox[0in][r]{\raisebox{3.2in}{\em b) \hspace{2.1in}}}
\vspace{-0.6in}
\begin{center}
{\small Figure~1: Mass distributions of the
a) $J/\psi$ and b) $\psi(2S)$. The mass cuts are indicated.}
\end{center}
\end{figure}

The $K^+$, $K^0_S$, and $K^*(892)$ candidates
must have $p_{T} > 1.25$ GeV/c in order to be combined with a $J/\psi$ or
$\psi(2S)$
to reconstruct a $B$ meson.  Individual cuts of $p_{T}(K) > 1.0$ and
$p_{T}(\pi) > 0.5$ GeV/c are also required for $K^*(892)^0$ candidates.

In the final $B$ reconstruction, all the decay tracks, except those from a
$K^0_S$, are vertex constrained, and the $J/\psi$ and $\psi (2S)$ candidates
are mass constrained to their world average values.
Any $B$ mesons with $p_{T} < 6.0$ GeV/c are rejected.
In the case of multiple candidates per event, only the one with the
best $\chi^2$ from
the constrained fit is kept.

The upper plots in Figure~2 show the invariant mass
distributions of all $B^+$ and $B^0$ candidates.
\begin{figure}[thb]
\epsfxsize=5.0in \epsfbox{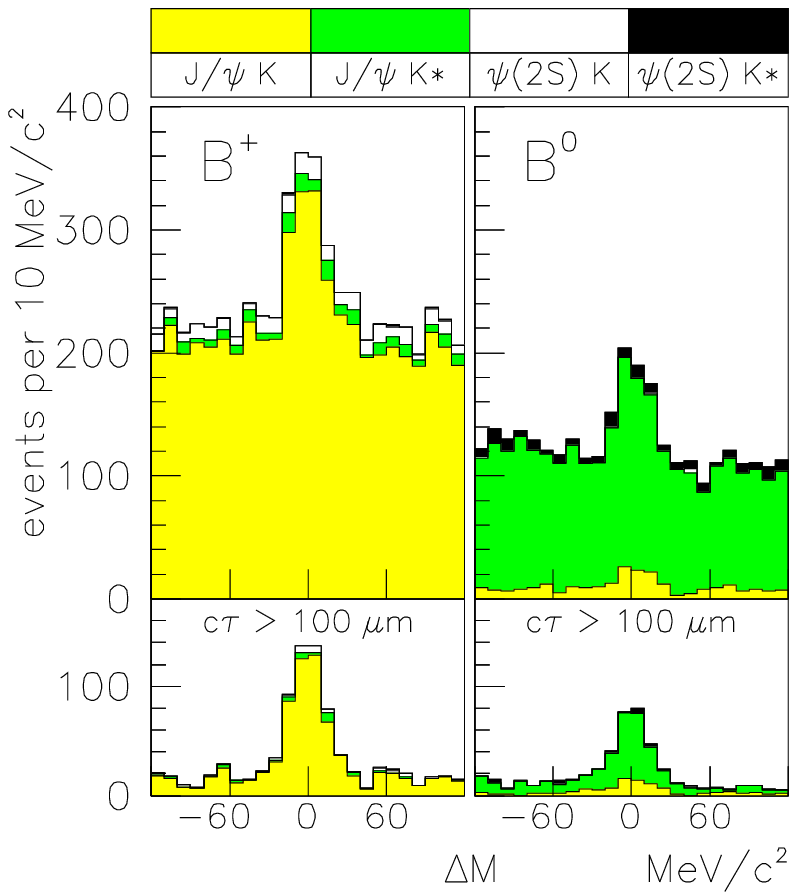}
\vspace{-0.2in}
\begin{center}
{\small Figure 2: Mass distributions of the fully reconstructed
$B$ samples, both with (upper) and without (lower) a $c\tau > 100~ \mu$m
cut. (CDF Preliminary)}
\end{center}
\label{lifemass}
\end{figure}
Background in these
distributions comes from combinations of $J/\psi$'s with tracks produced
during the $b$-quark fragmentation or with other remnants of the
$p \overline{p}$ collision.  These tracks should reconstruct to the primary
vertex and consequently the background is smallest for events where the
decay distance is large.  This is demonstrated in the lower plots of
Figure~2 where a $c\tau > 100 \: \mu$m cut is made to
illustrate this point.  The contributions from the different decay modes is
given by the shading and, as expected, the $B^+ \rightarrow J/\psi K^+$ decay
is dominant for charged $B$'s and the $B^0 \rightarrow J/\psi K^*(892)^0$
channel has the largest rate for neutral $B$'s.

For the lifetime analysis, we define the signal region to be
within $\pm 30$ MeV/c$^2$ of
the world average $B$ mass~\cite{pdg}.
Sideband regions are defined to be between 60 and 120 MeV/c$^2$
away from the
world average.
This selection excludes the mass region where $B$'s with a missing $\pi$
would be typically reconstructed.

The $B^+$ and $B^0$ proper
decay length distributions, for both the
signal
and sideband regions, are shown in Figure~3.
\begin{figure}[thb]
\epsfxsize=5.0in \epsfbox{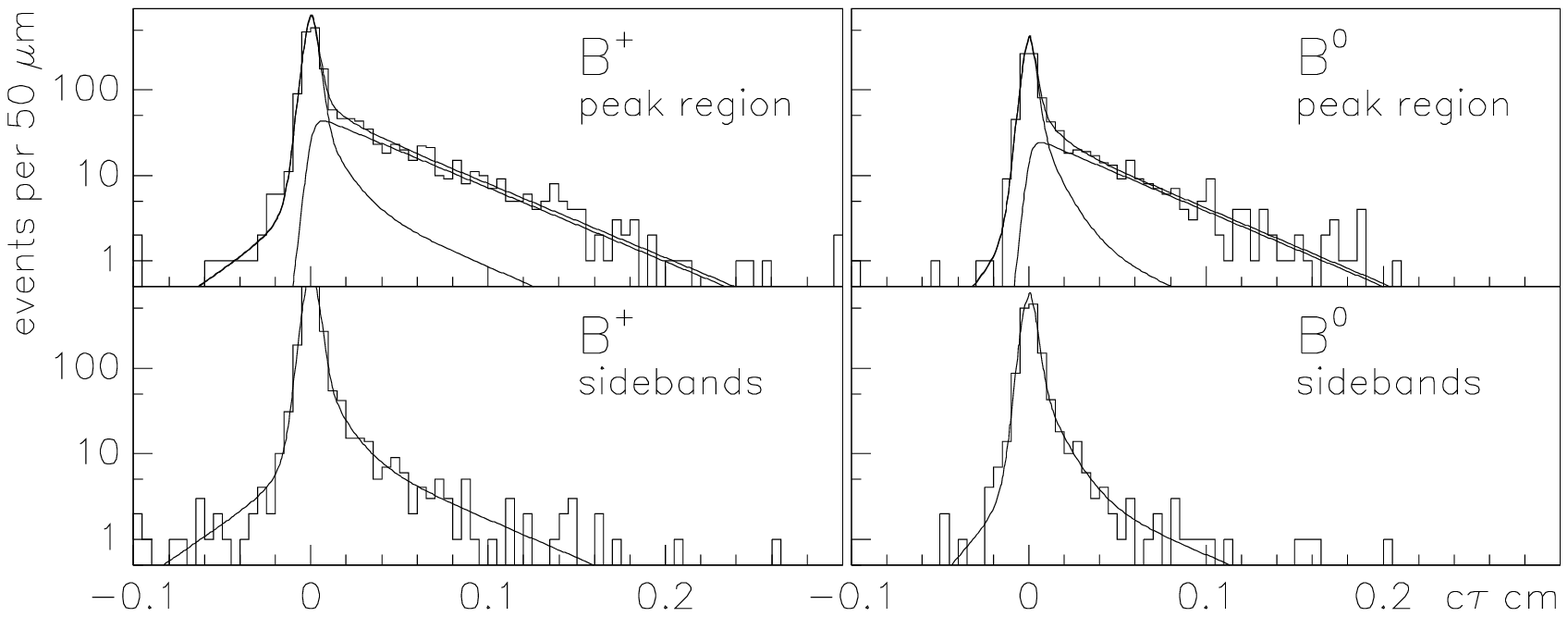}
\vspace{-0.25in}
\begin{center}
{\small Figure 3: The proper decay length ($c\tau$) distributions
of the fully reconstructed $B$ samples. The fits (curves)
are described in the text. (CDF Preliminary)}
\end{center}
\label{lifectau}
\end{figure}
The superimposed curves are the results of separate unbinned
likelihood fits
for the $B^+$ and $B^0$ lifetimes.
The signal region fit consists of a lifetime exponential
convoluted with a Gaussian resolution function, while the background is
modeled with a Gaussian plus
asymmetric exponential tails.
The signal and background distributions are fit simultaneously.
The fits indicate that there are $524 \pm 29$  charged and $285 \pm 21$
neutral $B$ mesons in the signal regions.  There is some residual positive
lifetime component to the background due to resolution effects in which
$B \rightarrow J/\psi K \pi^0$ decays are reconstructed in the $B$ mass
sideband region.  The results of the lifetime fits to the $c\tau$ distributions
are $c\tau^+ = 503 \pm 26 \: \mu$m and $c\tau^0 = 492 \pm 34 \: \mu$m.

Residual misalignment, trigger bias, and beam stability give the dominant
contributions to the systematic uncertainty.  However, these are common to
the $B^+$ and $B^0$ lifetime measurements and cancel in the lifetime ratio.
The other systematic contributions are significantly reduced compared to the
previous Run~1a analysis due to the increased stability of the $c\tau$
distributions from the larger statistics.  Table~1 gives the sources of
systematic uncertainty.

\begin{table}[htb]
\begin{center}
\begin{tabular}{|l|r|r|} \hline
Source of Uncertainty           & $B^+$  & $B^0$ \\ \hline \hline
Residual misalignment           & 10 $\mu$m & 10 $\mu$m \\
Trigger bias                    & 11 $\mu$m & 11 $\mu$m \\
Beam stability                  &  8 $\mu$m &  8 $\mu$m \\ \hline
Resolution (scale)              &  1 $\mu$m &  1 $\mu$m \\
Resolution (tails)              &  1 $\mu$m &  4 $\mu$m \\
Background shape                &  1 $\mu$m &  2 $\mu$m \\
Fitting procedure bias          &  2 $\mu$m &  1 $\mu$m \\ \hline \hline
Total                           & 17 $\mu$m & 18 $\mu$m \\
\hline
\end{tabular}
\end{center}
\caption{Summary of systematic uncertainties for the exclusive
$B^+$ and $B^0$ lifetime measurements.}
\end{table}

In the combined Run~1a + 1b data sample (67.7 pb$^{-1}$), the
preliminary measurements of $\tau^+$, $\tau^0$, and $\tau^+/\tau^0$ using
exclusive $B \rightarrow J/\psi K$ decays are:
\begin{displaymath}
\begin{array}[b]{rll}
\tau^+_{excl} &= & 1.68 \pm 0.09 \: {\rm (stat)} \pm 0.06 \: {\rm (syst)}
\: {\rm ps} \\
\tau^0_{excl} &= & 1.64 \pm 0.11 \: {\rm (stat)} \pm 0.06 \: {\rm (syst)}
\: {\rm ps} \\
(\tau^+/\tau^0)_{excl} &= & 1.02 \pm 0.09 \: {\rm (stat)} \pm 0.01
\: {\rm (syst)}
\end{array}
\end{displaymath}

We see that the uncertainty on the $B^+$ ($B^0$) lifetime is only
6.4\% (7.6\%) and the precision on the lifetime ratio is dominated by
statistics and is less than 10\%.

\begin{figure}[thb]
\hspace{0.05in}
\parbox[b]{3.0in}{\epsfxsize=2.9in\epsfbox{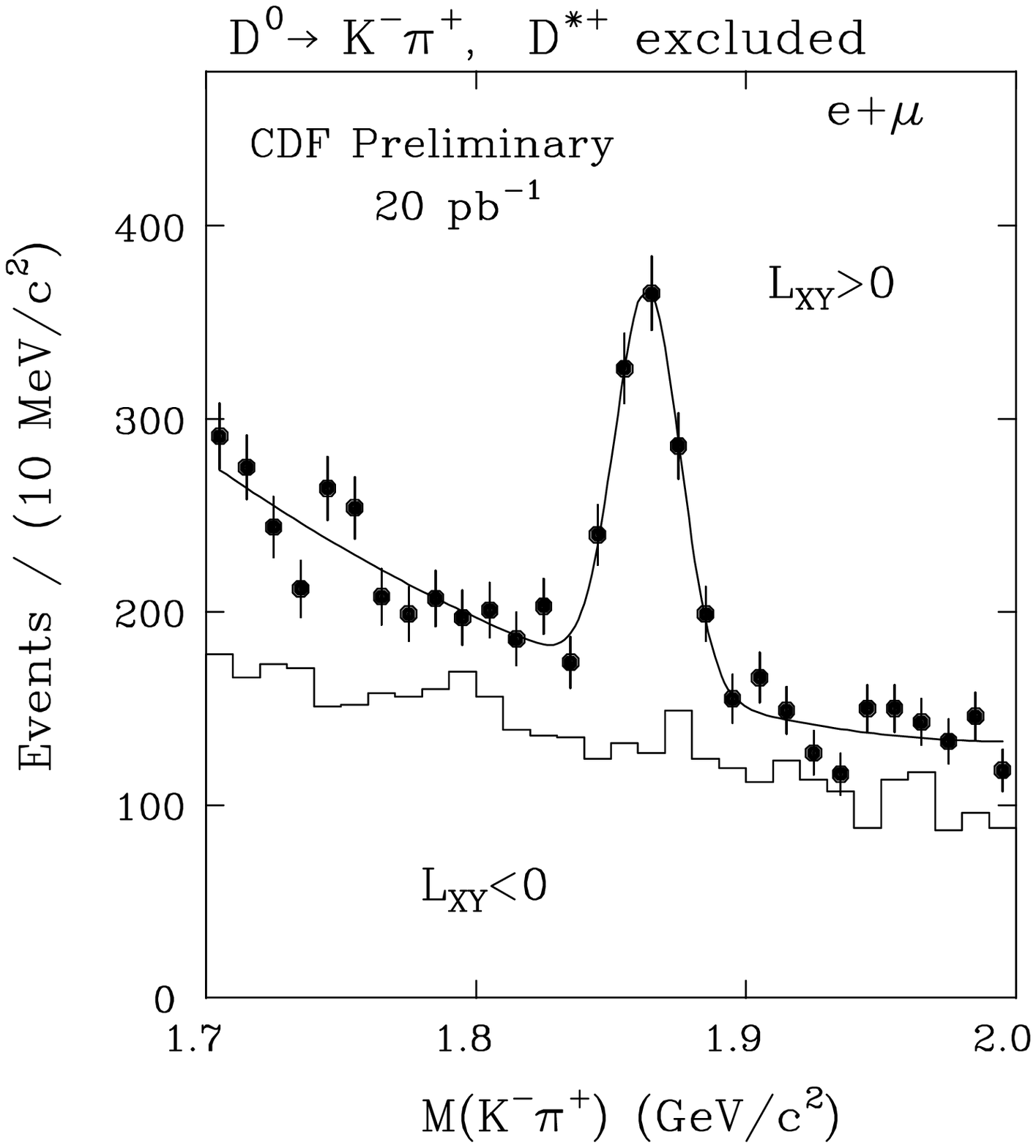}}%
\makebox[0in][r]{\raisebox{3.1in}{\em  a) \hspace{2.1in}}}
\hfill
\parbox[b]{3.0in}{\epsfxsize=2.9in\epsfbox{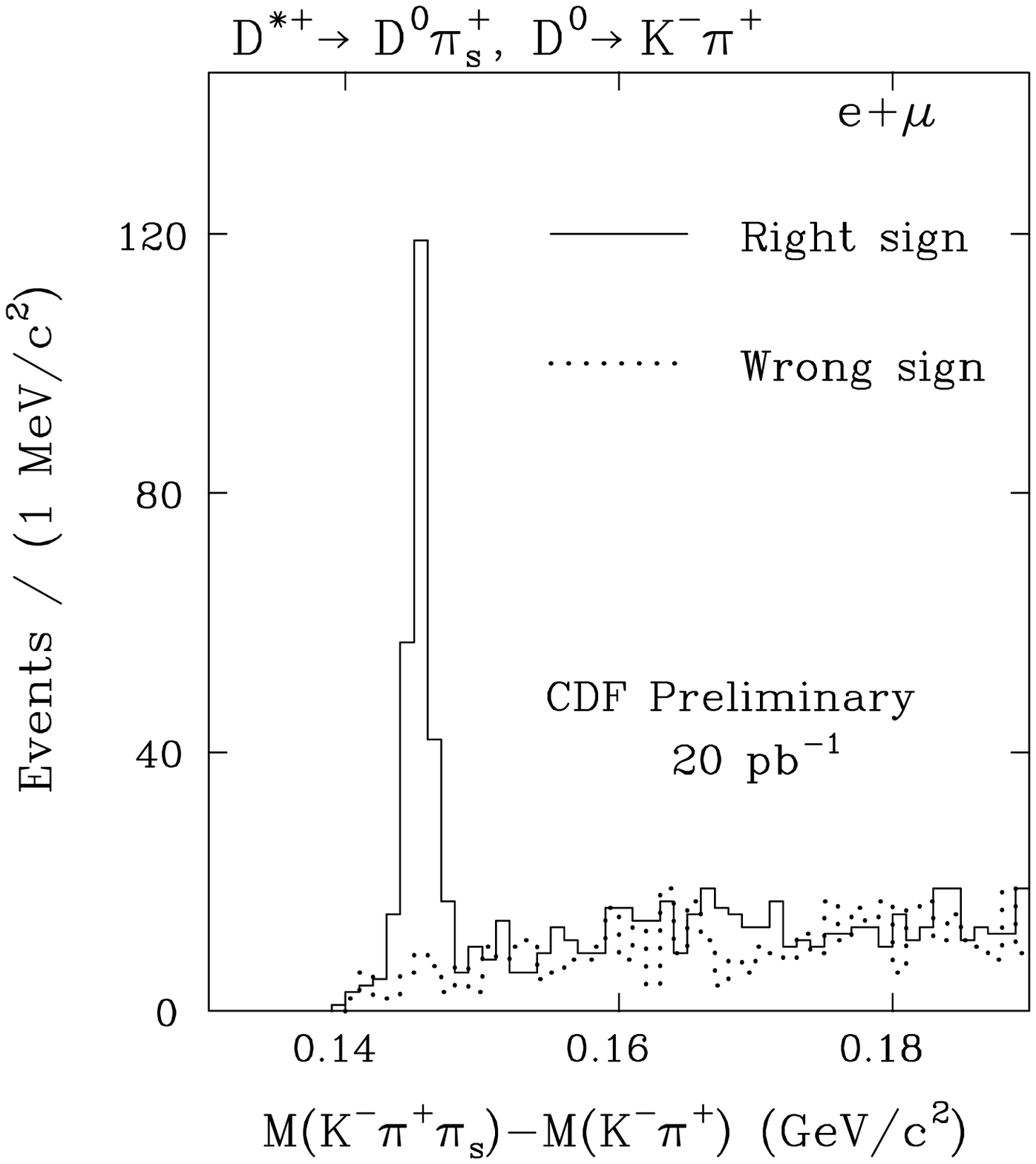}}%
\makebox[0in][r]{\raisebox{3.1in}{\em b) \hspace{2.1in}}}
\vspace{-0.8in}
\begin{center}
{\small Figure~4: a) The $K^- \pi^+$ invariant mass distribution.
Events from $D^{*+}$ decay are excluded.
b) The distribution of the mass difference, $\Delta m = m(K^- \pi^+ \pi_s) -
m(K^- \pi^+)$, for $D^{*+} \rightarrow D^0 \pi^+_s,
D^0 \rightarrow K^- \pi^+$ candidates.}
\end{center}
\label{bfig1}
\end{figure}

We now turn to a measurement of the charged and neutral
$B$ meson lifetimes using semileptonic decays, as has been
previously done
by several LEP
experiments~\cite{blife_LEP}.  Partially
reconstructed semileptonic decays of $B$ mesons, namely a lepton in
association with a $D^{0}$ or $D^{*+}$ meson will provide {\it nearly
orthogonal} samples of charged and neutral $B$ mesons and thus enable a
determination of their individual lifetimes.
\begin{figure}[htb]
\parbox[b]{3.0in}{\epsfxsize=3.0in\epsfbox{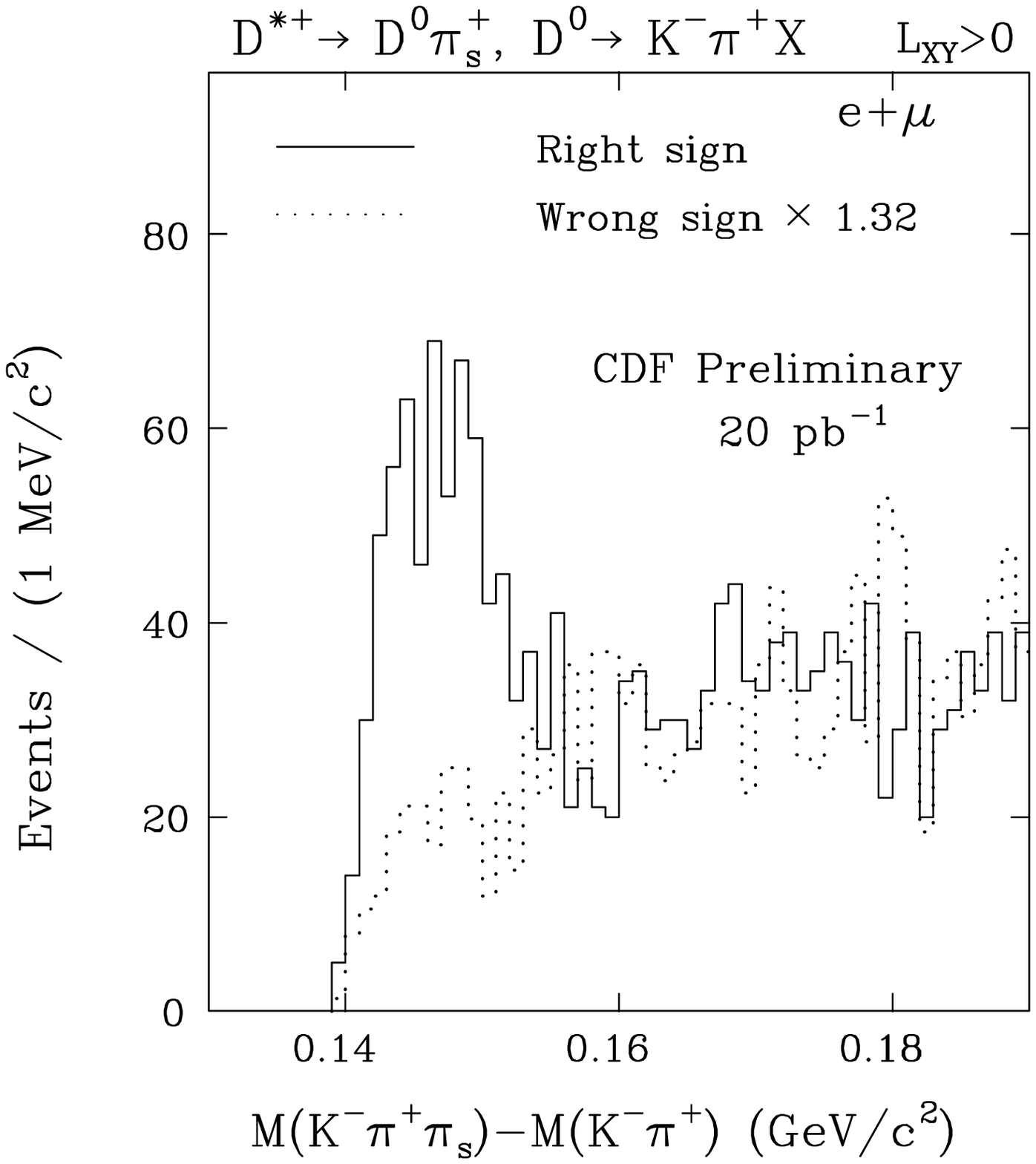}}%
\makebox[0in][r]{\raisebox{3.1in}{\em  a) \hspace{2.1in}}}
\hfill
\parbox[b]{3.0in}{\epsfxsize=3.0in\epsfbox{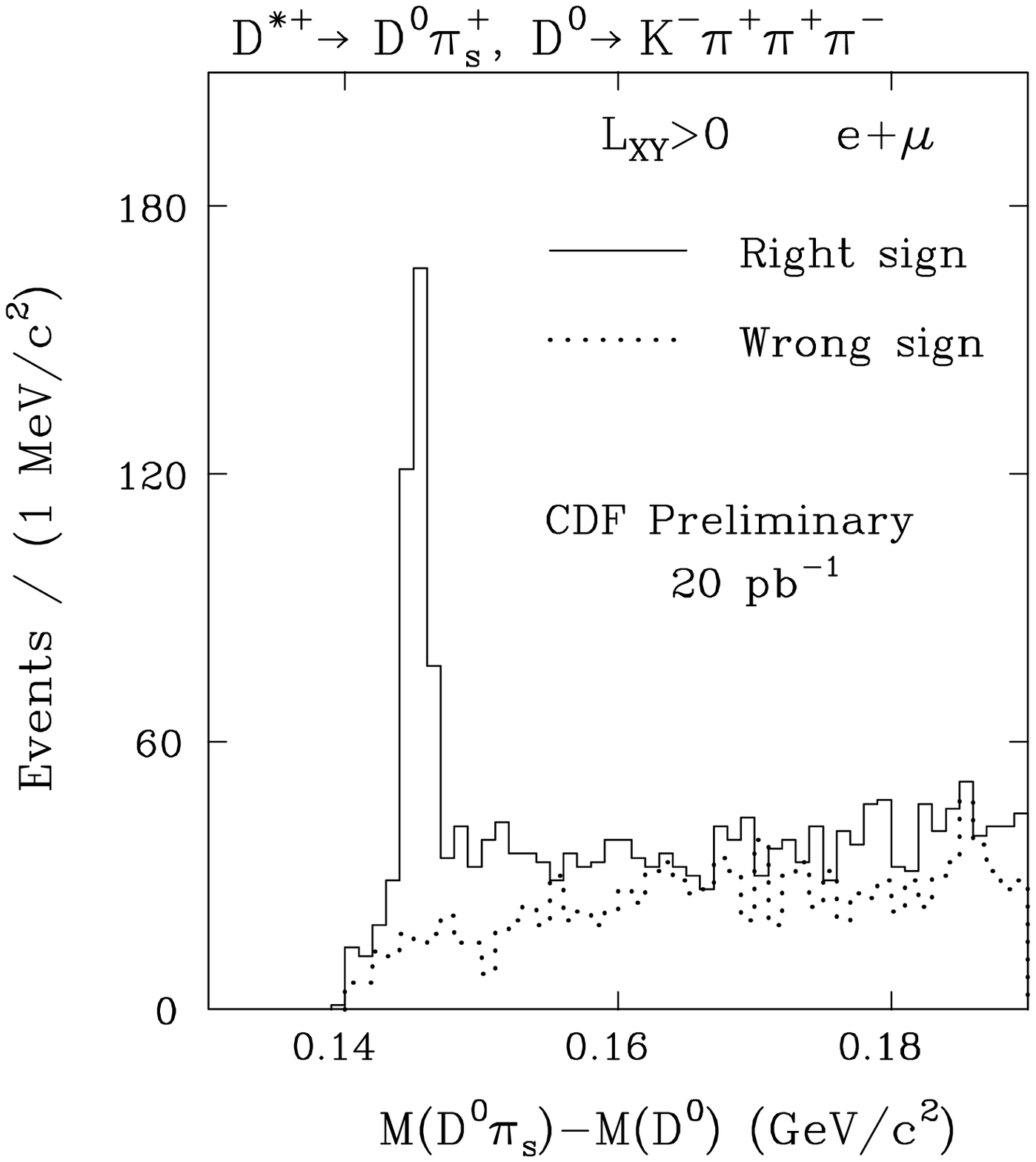}}%
\makebox[0in][r]{\raisebox{3.1in}{\em b) \hspace{2.1in}}}
\vspace{-0.8in}
\begin{center}
{\small Figure~5: The mass difference distribution for
a) $D^{*+} \rightarrow D^0 \pi^+_s, D^0 \rightarrow K^- \pi^+ X$ (satellite)
and b) $D^{*+} \rightarrow D^0 \pi^+_s, D^0 \rightarrow K^- \pi^+ \pi^+ \pi^-$
candidates.}
\end{center}
\label{bfig2}
\end{figure}

This new preliminary measurement uses both the single electron and muon samples
in the $\sim 20$ pb$^{-1}$ Run~1a data.
Four sources of $D^0$ mesons are considered in this
analysis: 1) $B^- \rightarrow l^- \overline{\nu} D^0 X,
D^0 \rightarrow K^- \pi^+$, where the $D^0$ is {\it not} from $D^{*+}
\rightarrow D^0 \pi^+_s$ (``$D^{0}$ sample''); 2) $\overline{B}^0
\rightarrow l^- \overline{\nu} D^{*+} X, D^{*+} \rightarrow D^0 \pi^+_s,
D^0 \rightarrow K^- \pi^+$ (``$D^{*+}, D^0 \rightarrow K^- \pi^+$ sample'');
3) $\overline{B}^0 \rightarrow l^- \overline{\nu} D^{*+} X,
D^{*+} \rightarrow D^0 \pi^+_s, D^0 \rightarrow K^- \pi^+ X$
(``satellite sample''); and 4) $\overline{B}^0
\rightarrow l^- \overline{\nu} D^{*+} X, D^{*+} \rightarrow D^0 \pi^+_s,
D^0 \rightarrow K^- \pi^+ \pi^+ \pi^-$
(``$D^{*+}, D^0 \rightarrow K^- \pi^+ \pi^+ \pi^-$ sample'').
The decay length of the $D^0$ projected onto the lepton-charm momentum
in the plane transverse to the colliding beams, $L_{XY}$, must satisfy
$L_{XY} > 0$ for the $D^{0}$ and satellite samples.
Figure~4a shows the resulting $K^- \pi^+$ mass distribution,
containing $560 \pm 40$ events in the $D^0$ peak, for the $D^{0}$ sample.
Figure~4b gives the mass difference distribution for the
$D^{*+}, D^0 \rightarrow K^- \pi^+$ sample, while Figures~5a and 5b give
the $\Delta m$ distributions for
the satellite sample and $D^{*+}, D^0 \rightarrow K^- \pi^+ \pi^+ \pi^-$
samples, respectively.
These figures show the production of $l-D^0$ combinations in the expected
(``right sign'') charge combinations and little evidence of any signal
above combinatoric
background in the ``wrong sign'' combinations.

	The lepton and $D^0$ tracks are intersected to determine the
$B$ decay vertex position and decay length from the primary vertex.  Since
the $B$ is only partially reconstructed, the $l-D^0$ system transverse
momentum can be used to determine a ``pseudo-$c\tau$'' value
$c\tau^* = L_B m_B/p_{T}(l + D^0) = c\tau/K$,
where $K$ is a momentum correction factor determined from Monte Carlo.  It is
determined separately for each $D^0$ signal sample.

	The signal $c \tau^*$ distributions are fit with an exponential
lifetime term convoluted with a Gaussian resolution function and the
momentum correction distribution.
The lifetime of the background in the signal region is determined from the
wrong sign and signal sideband distributions and is modeled by
a Gaussian resolution function plus exponential tails.
Figure~6 shows the results of the lifetime fits in the $D^{0}$ and satellite
signal samples.
\begin{figure}[htb]
\parbox[b]{3.0in}{\epsfxsize=3.0in\epsfbox{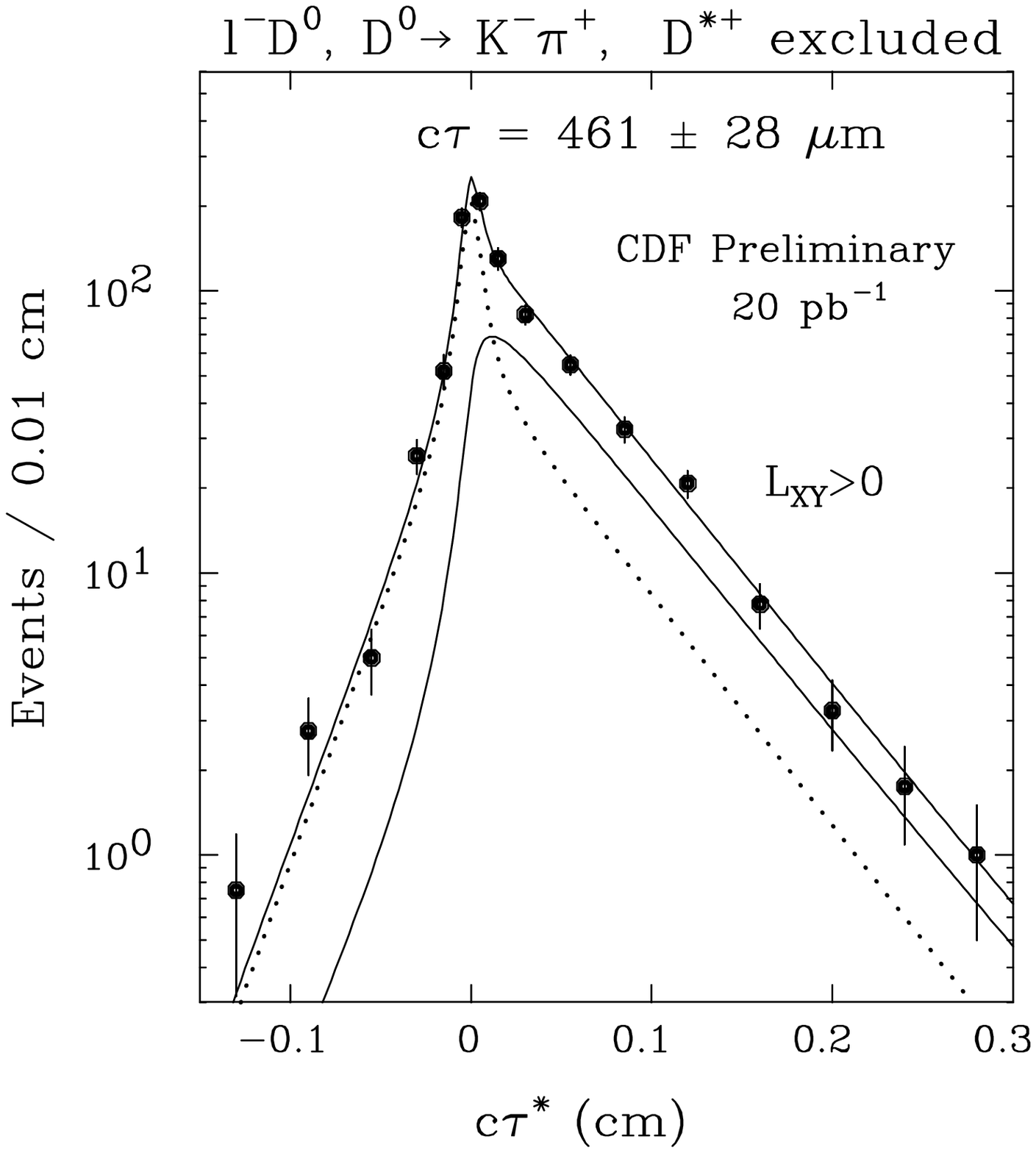}}%
\makebox[0in][r]{\raisebox{3.1in}{\em  a) \hspace{2.1in}}}
\hfill
\parbox[b]{3.0in}{\epsfxsize=3.0in\epsfbox{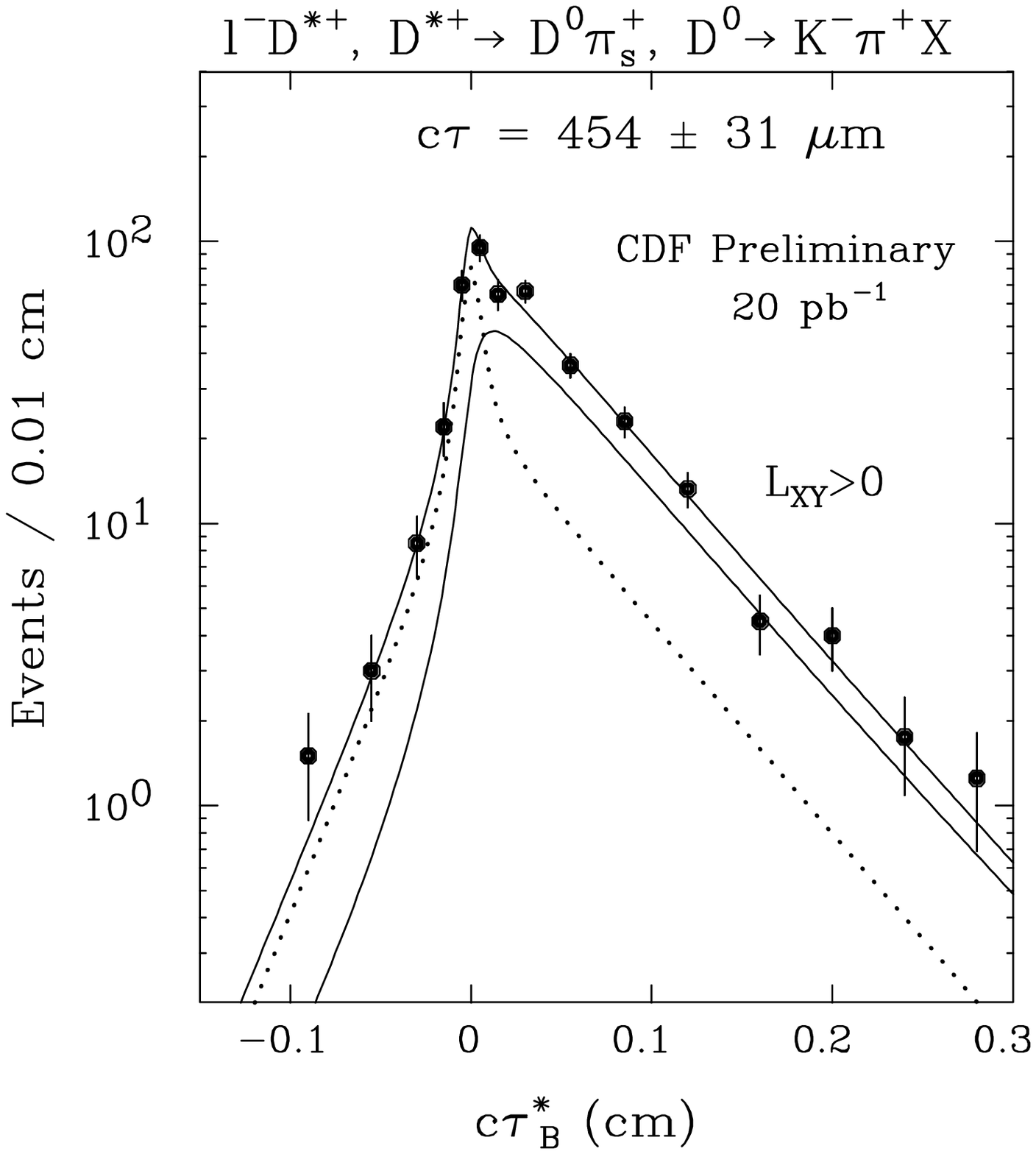}}%
\makebox[0in][r]{\raisebox{3.1in}{\em b) \hspace{2.1in}}}
\vspace{-0.8in}
\begin{center}
{\small Figure~6: Distribution of pseudo-$c\tau$ for the
a) $D^{0}$ and b) satellite samples.}
\end{center}
\label{bfig3}
\end{figure}
The fit quality and results for the other $D^{*+}$ samples are similar.
The fraction of $B^{-}$ and $\overline{B^{0}}$ contributing to each of
the $D^{0}, D^{*+}$, and satellite samples is  determined and includes the
effects of cross-talk due to:
1) the $\pi^{+}_{s}$ reconstruction efficiency, $\epsilon(\pi^{+}_{s})
= 0.93^{+0.07}_{-0.21}$.  A missed spectator pion from $D^{*+}$ decay can
cause a
$D^{0}$ to be associated with $B^-$ rather than $\overline{B}^0$;
2) the $D^{**}$ fraction, $f^{**} = {\rm BR}(\overline{B}
\rightarrow l^- \overline{\nu} D^{**})/{\rm BR}(\overline{B}
\rightarrow l^- \overline{\nu} X) = 0.36 \pm 0.12;$~\cite{CLEO}
3) the fraction of $D^{**}$ decaying to $D^{*}$, from Monte Carlo simulation
is found to be
${\rm BR}(D^{**} \rightarrow D^* \pi)/
({\rm BR}(D^{**} \rightarrow D^* \pi) + {\rm BR}(D^{**} \rightarrow D \pi))
= 0.78$; and 4)
the charged-to-neutral lifetime ratio can affect the event mixture,
${\rm BR}(B^- \rightarrow l^- \overline{\nu} X)/{\rm BR}(\overline{B^{0}}
\rightarrow l^- \overline{\nu} X) = \tau(B^-)/\tau(\overline{B^0})$.
In spite of these effects, we find that the $D^{0}$ and $D^{*+}$ signals
provide {\it nearly orthogonal} samples of $B^-$ and $\overline{B^0}$ mesons.
A combined likelihood function is used to simultaneously fit the
signal samples for the $B^-$ and $\overline{B^0}$ meson lifetimes.
Variations in the sample composition due to the above effects are included in
the systematic uncertainty.  The results are:
\begin{displaymath}
\begin{array}[b]{rll}
\tau^{+}_{semi} &= & 1.51 \pm 0.12 \: {\rm (stat)} \pm 0.08 \:
{\rm (syst)} \:
{\rm ps} \\
\tau^{0}_{semi} &= & 1.57 \pm 0.08 \: {\rm (stat)} \pm 0.07 \:
{\rm (syst)} \: {\rm ps} \\
(\tau^{+}/\tau^{0})_{semi} & = & 0.96 \pm 0.10 \: {\rm (stat)} \pm 0.05
\: {\rm (syst)}
\end{array}
\end{displaymath}

We can separate out the relatively small correlations between the
exclusive and semileptonic mode lifetime measurements due to residual
misalignment and beam stability and compute CDF average values for
$\tau^{+}$, $\tau^{0}$, and $\tau^{+}/\tau^{0}$.  We find:
\begin{displaymath}
\begin{array}[b]{rll}
\tau^{+}_{CDF} &= & 1.62 \pm 0.09 \: {\rm ps} \\
\tau^{0}_{CDF} &= & 1.60 \pm 0.09 \: {\rm ps} \\
(\tau^{+}/\tau^{0})_{CDF} & = & 1.00 \pm 0.07
\end{array}
\end{displaymath}

These can be compared to the latest LEP $B$ lifetime averages~\cite{morave}:
\begin{displaymath}
\begin{array}[b]{rll}
\tau^{+}_{LEP} &= & 1.68 \pm 0.07 \: {\rm ps} \\
\tau^{0}_{LEP} &= & 1.56 \pm 0.07 \: {\rm ps} \\
(\tau^{+}/\tau^{0})_{LEP} & = & 1.08 \pm 0.08
\end{array}
\end{displaymath}

These results are in good agreement and the precision on the CDF lifetime
averages is now nearly identical to the latest LEP values.

\begin{figure}[thb]
\parbox[b]{2.8in}{\epsfxsize=2.8in\epsfbox{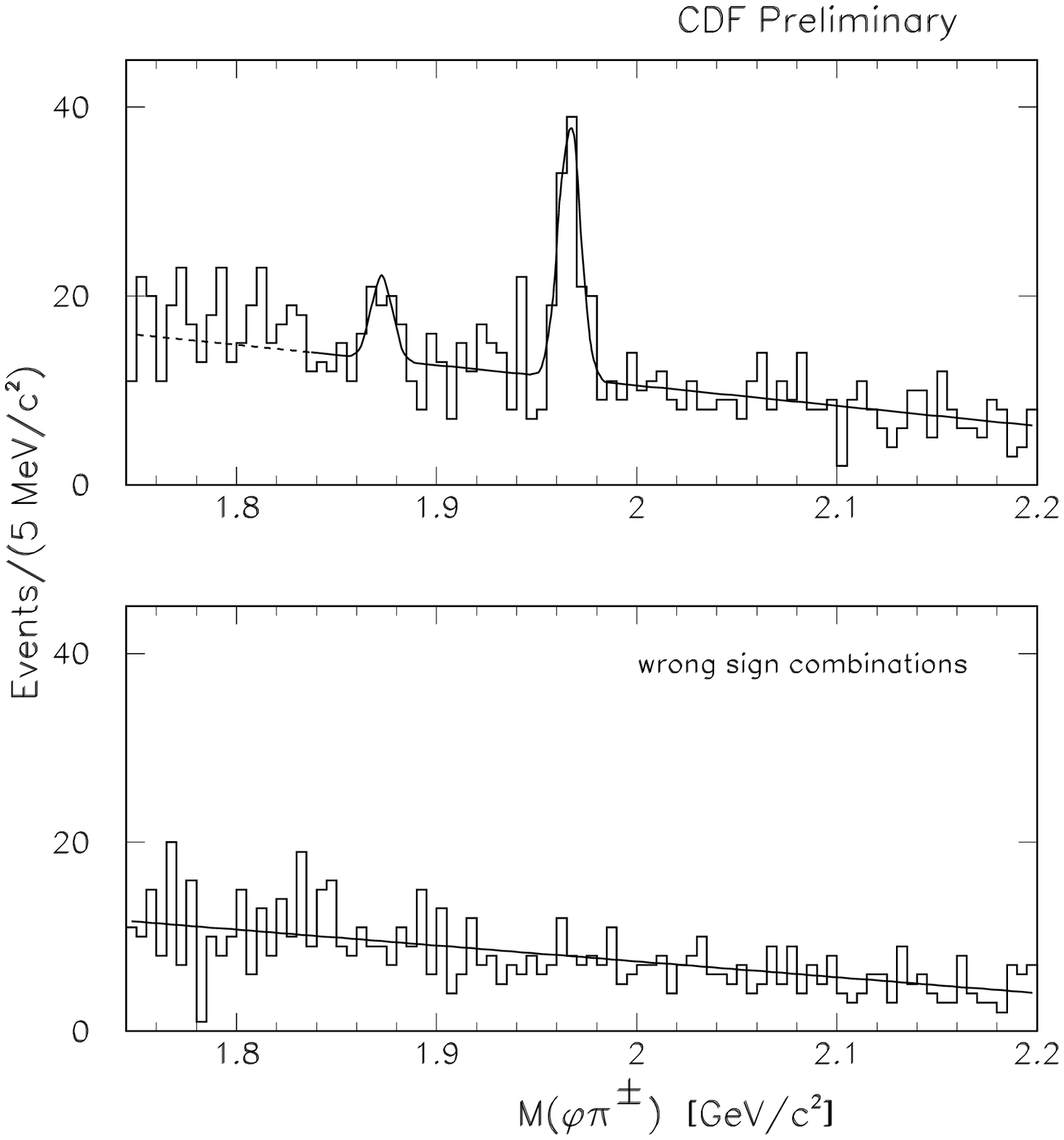}}%
\makebox[0in][r]{\raisebox{3.2in}{\em  a) \hspace{2.3in}}}
\hfill
\parbox[b]{2.8in}{\epsfxsize=2.8in\epsfbox{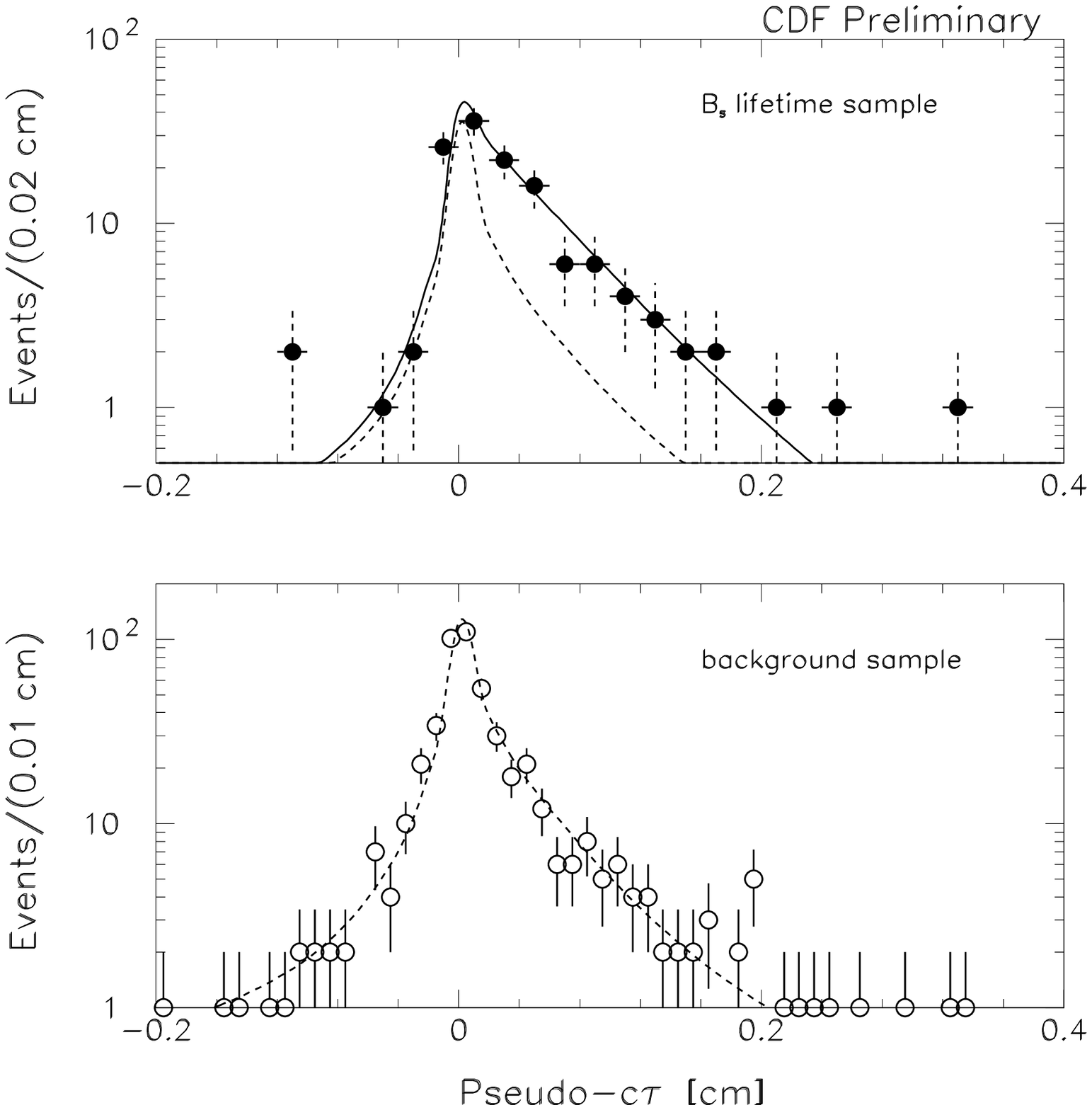}}%
\makebox[0in][r]{\raisebox{3.2in}{\em b) \hspace{2.3in}}}
\vspace{-0.6in}
\begin{center}
{\small Figure~7: a) The $\phi\pi^{-}$ mass distribution for
``right sign'' and ``wrong sign''  lepton-$D_s$ combinations.
b) Pseudo-$c\tau$ distribution of the $l^{-}D^{+}_{s}$ signal
sample showing the lifetime fits of the combined (signal plus background) and
background distributions separately.}
\end{center}
\label{bsfig1}
\end{figure}
\section{$B_s$ Meson Lifetime}
A similar technique has been used to measure the $B_s$ meson lifetime using
semileptonic $B_s \rightarrow l \overline{\nu} D_s, D_s \rightarrow \phi \pi,
\phi \rightarrow K^+ K^-$ decays~\cite{cdf_bslife}.  Precision measurement
of the $B_s$ meson lifetime is particularly important since it has been
suggested by recent theory calculations that the lifetime between the two
CP eigenstates produced by the mixing of the $B_s$ and $\overline{B}_s$ may
be different by as much as 20\%~\cite{isi}.  Such an effect should manifest
itself as a
difference in lifetimes between the $B_s$ semileptonic decay, which is
an equal mixture of the two CP states, and the decay
$B_s \rightarrow J/\psi \phi$, which is expected to be dominated by the
CP even state.
Figure~7a shows the $K^+ K^- \pi^+$ invariant mass spectrum after
all cuts for the combined electron and muon samples.  Some $76 \pm 8$ events
are found in the right-sign mass peak and a hint of the Cabbibo suppressed
$D^+ \rightarrow \phi \pi^+$ decay is seen.  Following the same procedure as
above, the $B_s$ lifetime from semileptonic $B_s \rightarrow
l \overline{\nu} D_s$ is measured to be (Figure~7b):
\begin{displaymath}
\begin{array}[b]{rll}
\tau_{s}^{semi} & = & 1.42 ^{+0.27}_{-0.23} \: {\rm (stat)} \pm 0.11
\: {\rm (syst)} \: {\rm ps}
\end{array}
\end{displaymath}
Finally, there is a new low-statistics measurement of the $B_s$ lifetime using
fully reconstructed $B_s \rightarrow J/\psi \phi, J/\psi
\rightarrow \mu^+ \mu^-,
\phi \rightarrow K^+ K^-$ decays~\cite{cdf_bslife}.  At least two of the four
daughter tracks are
required to be reconstructed in the SVX.  Based on a sample of 10 events, the
$B_s$ lifetime using exclusive $B_s \rightarrow J/\psi \phi$ is measured to be:
\begin{displaymath}
\begin{array}[b]{rll}
\tau_s^{excl} & = & 1.74 ^{+1.08}_{-0.69} \: {\rm (stat)} \pm 0.07 \:
{\rm (syst)} \: {\rm ps}
\end{array}
\end{displaymath}
Both of these measurements will greatly benefit from the addition of the
Run~1b data and a first measurement of the $B_{s,long} - B_{s,short}$
lifetime difference should be possible.
\section{$B$ Meson Mass Measurements}
	The CDF $B^+$, $B^0$, and $B_s$ mass measurements in the Run~1a data
were reported previously~\cite{lewis}.  The work continues to
evaluate the tracking systematics in the current Run~1b data.  Sizeable
samples of fully reconstructed $B^+ \rightarrow J/\psi K^+$,
$B^0 \rightarrow J/\psi K^*(892)^0$, $B^0 \rightarrow J/\psi K^0_S$, and
$B_s \rightarrow J/\psi \phi$ decays will be available for precision $B$
mass measurements in the combined Run~1a + 1b data sample.  Figures~8a, 8b,
and 9a give the reconstructed mass peaks currently
under analysis after standard selection cuts.
\begin{figure}[thb]
\parbox[b]{3.0in}{\epsfxsize=2.9in\epsfbox{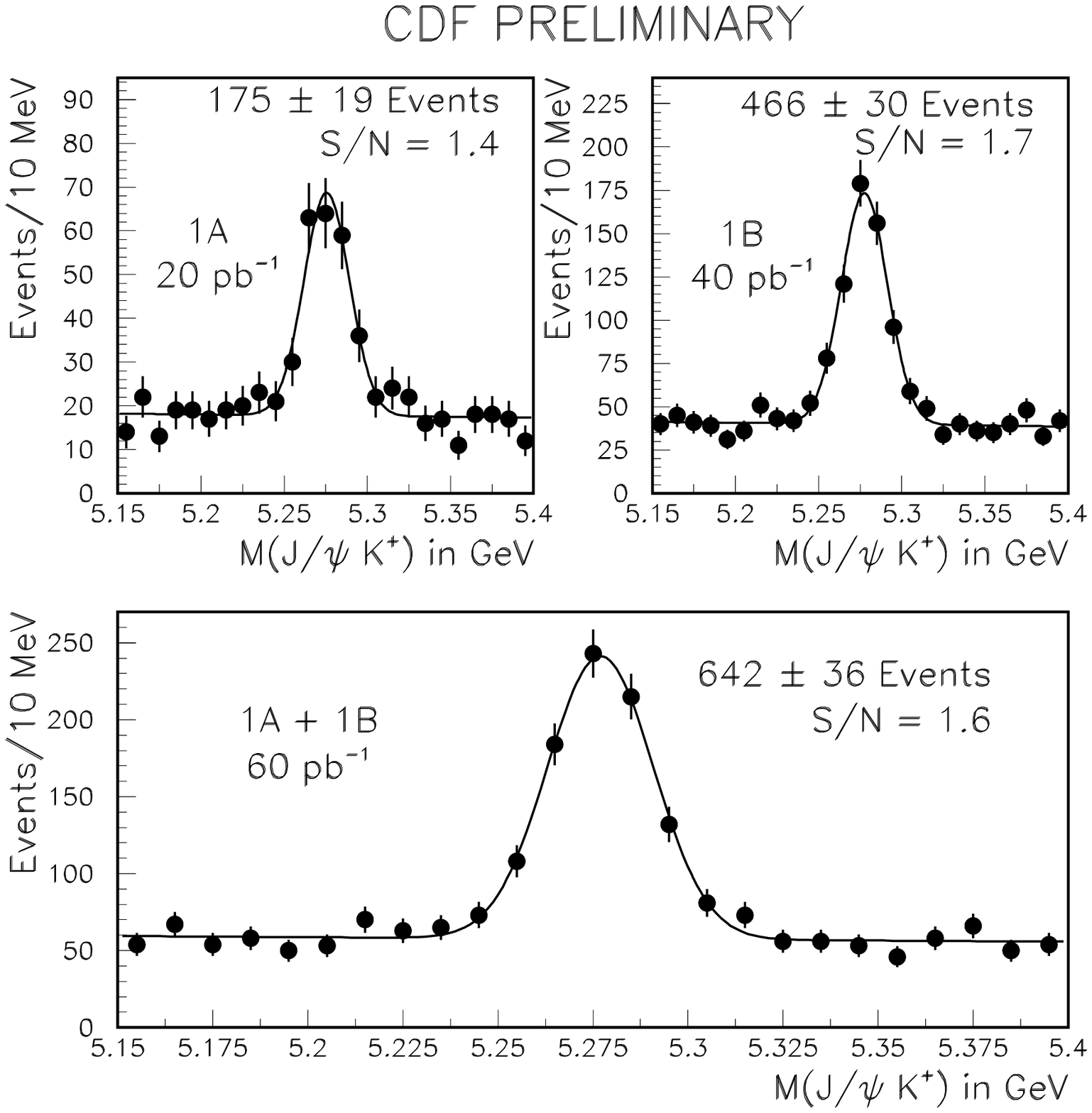}}%
\makebox[0in][r]{\raisebox{3.65in}{\em  a) \hspace{2.4in}}}
\hfill
\parbox[b]{3.0in}{\epsfxsize=2.9in\epsfbox{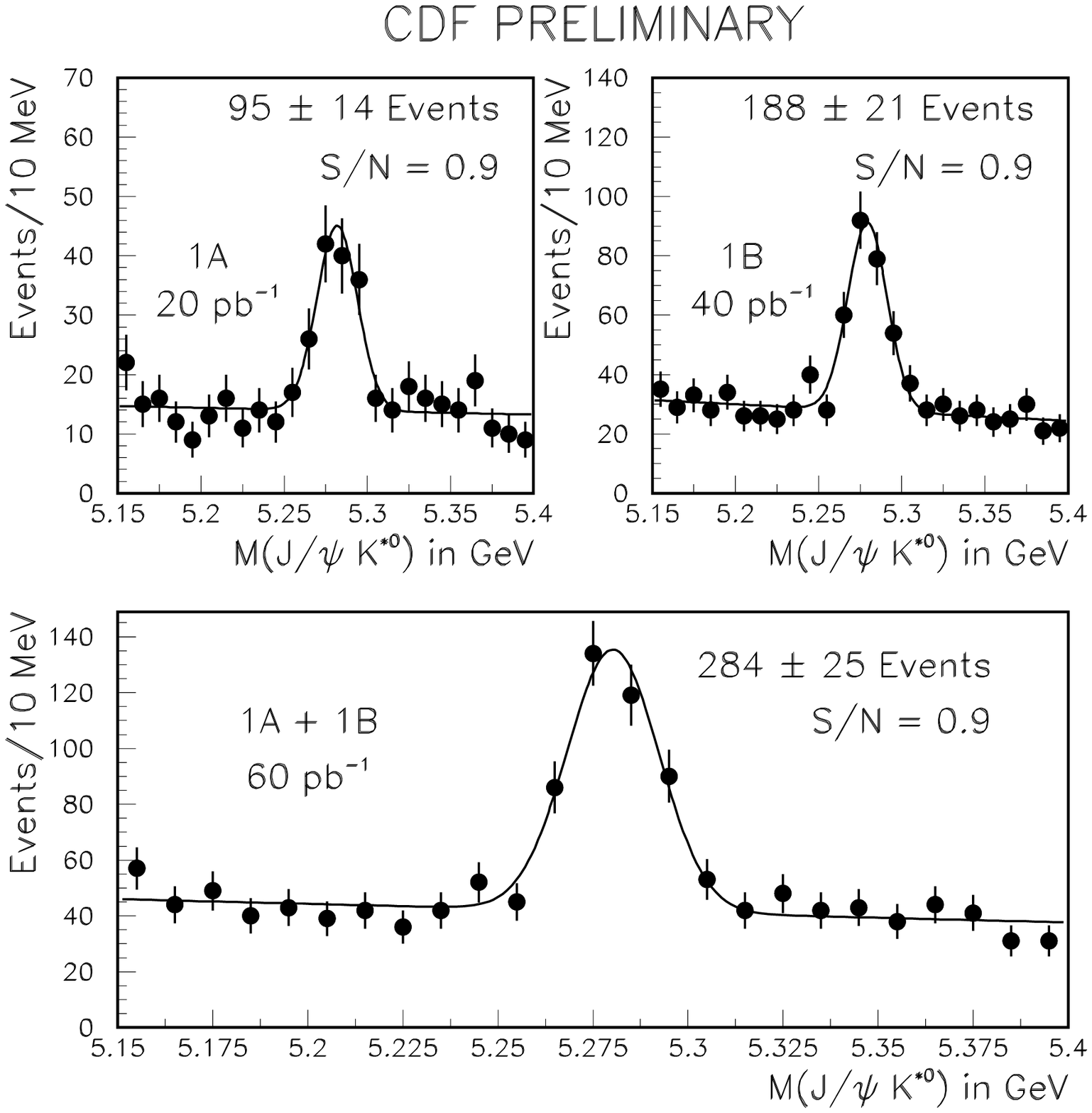}}%
\makebox[0in][r]{\raisebox{3.65in}{\em b) \hspace{2.4in}}}
\vspace{-0.6in}
\begin{center}
{\small Figure~8: Reconstructed a) $B^+ \rightarrow J/\psi K^+$
and b) $B^0 \rightarrow J/\psi K^*(892)^0$ mass
distributions in the 20 pb$^{-1}$ Run~1a, 40 pb$^{-1}$ Run~1b, and
combined 60 pb$^{-1}$ Run~1a + 1b data samples.}
\end{center}
\end{figure}

\section{Ratio of Branching Ratios}

Using its large sample of $J/\psi \rightarrow \mu^+ \mu^-$ decays, CDF has
measured the ratio of branching ratios for many exclusive $B$ decay modes
involving a $J/\psi$ in the final state.  Table~2 lists the latest
preliminary measurements.  Of particular interest is the updated
$BR(B_{d} \rightarrow J/\psi K^{0})/BR(B_{u} \rightarrow J/\psi K^{+})$
result and a new observation of Cabbibo-suppressed
$B^+ \rightarrow J/\psi \pi^+$ decays (Figure~9b) and
a measurement of
$BR(B_u \rightarrow J/\psi \pi^+)/BR(B_{u} \rightarrow J/\psi K^{+})$.
This latter result is in good agreement with theoretical expectations and the
recent CLEO measurement of $(4.3 \pm 2.3)\%$.~\cite{cleo1}

\begin{figure}[thb]
\parbox[b]{3.0in}{\epsfxsize=2.8in\epsfbox{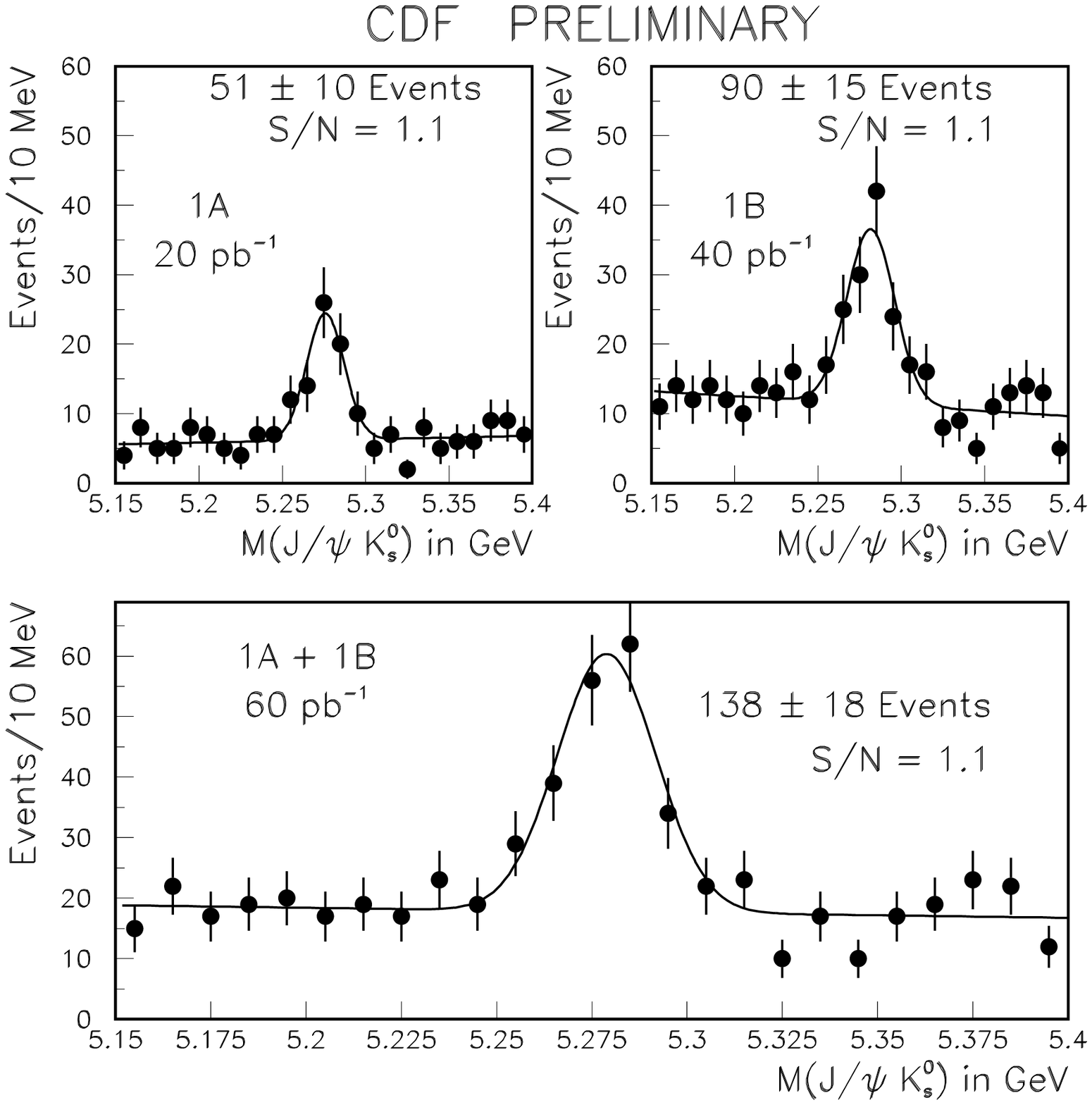}}%
\makebox[0in][r]{\raisebox{3.5in}{\em  a) \hspace{2.4in}}}
\hfill
\parbox[t]{3.0in}{\epsfxsize=3.2in\epsfbox{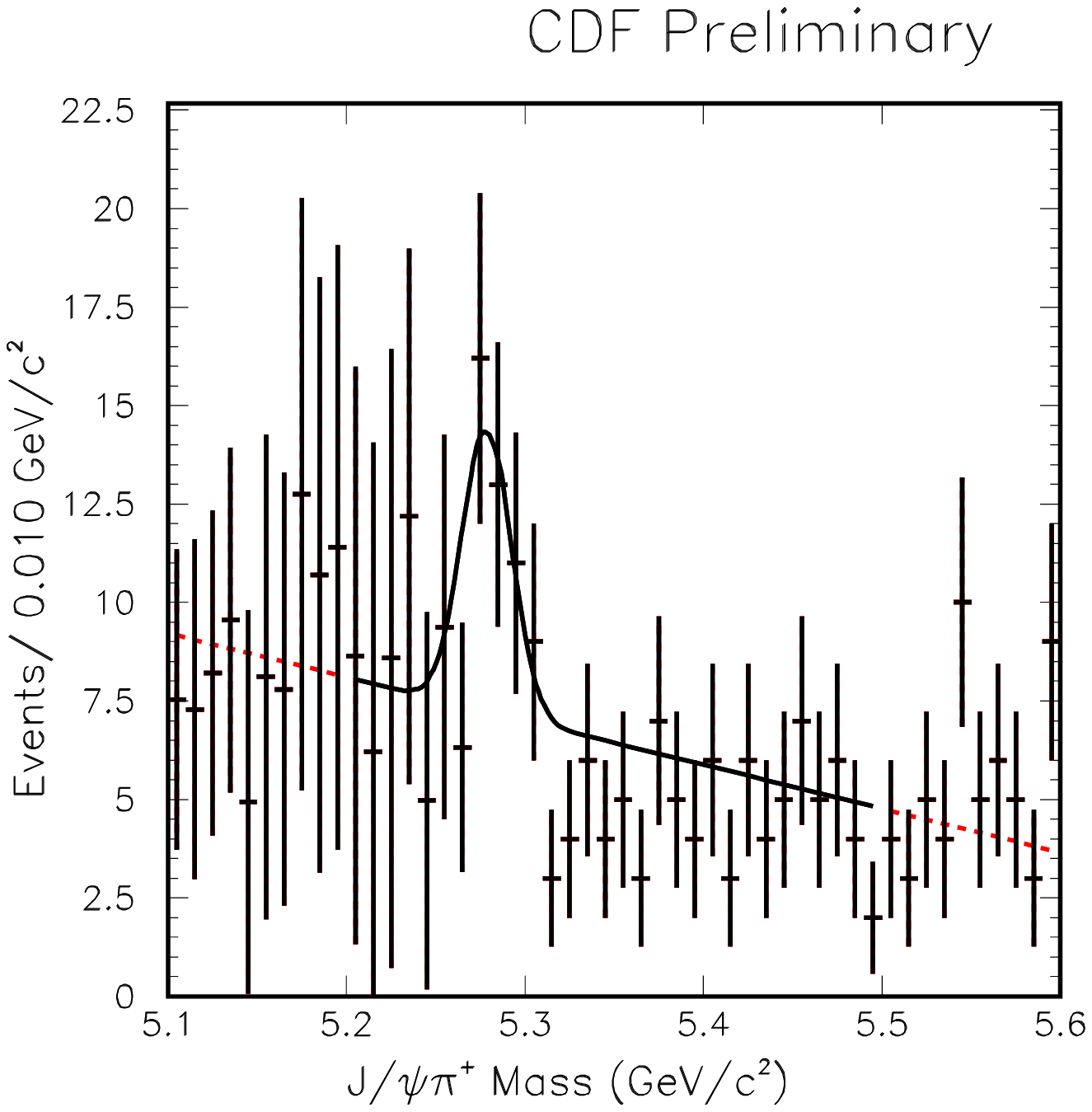}}%
\makebox[0in][r]{\raisebox{3.55in}{\em b) \hspace{2.3in}}}
\vspace{-0.6in}
\begin{center}
{\small Figure 9: a) Reconstructed $B^0 \rightarrow J/\psi K^0_S$ mass
distributions in the 20 pb$^{-1}$ Run~1a, 40 pb$^{-1}$ Run~1b, and
combined 60 pb$^{-1}$ Run~1a + 1b data samples. b) Reconstructed
$B^+ \rightarrow J/\psi \pi^+$ mass distribution ($25.1 \pm 8.4$ events are
observed in $\sim 65$ pb$^{-1}$).}
\end{center}
\end{figure}

The $b \rightarrow {\rm hadron}$ fractions and their errors remain to
be determined at the Tevatron.
However, if we assume that
$\frac{f(b \rightarrow B_{d})}{f(b \rightarrow B_{u})} =
\frac{0.375}{0.375} = 1.0$ and
$\frac{f(b \rightarrow B_{s})}{f(b \rightarrow B_{d})} = \frac{0.15}{0.375} =
0.40$, then we obtain the results in the lower part of Table~2.  The additional
Run~1b data will significantly improve the precision of the ratio of
branching ratio results.

\begin{table}[hbt]
\begin{center}
{\small Table~2. Preliminary CDF Ratio of Branching Ratio Results.}
\begin{tabular}{|c|c|} \hline
{\boldmath $BR$} Ratio & Measured Value \\ \hline
$\frac{f(b \rightarrow B_{d})}{f(b \rightarrow B_{u})} \times
\frac{BR(B_{d} \rightarrow J/\psi K^{0})}{BR(B_{u} \rightarrow J/\psi K^{+})}$
& $1.14 \pm 0.23 {\rm (stat)} \pm 0.08 {\rm (syst)}$ \\ \hline
$\frac{f(b \rightarrow B_{d})}{f(b \rightarrow B_{u})} \times
\frac{BR(B_{d} \rightarrow J/\psi K^{*}(892)^{0})}{BR(B_{u} \rightarrow
J/\psi K^{+})}$ & $1.68 \pm 0.34 {\rm (stat)} \pm 0.22 {\rm (syst)}$ \\ \hline
$\frac{f(b \rightarrow B_{s})}{f(b \rightarrow B_{u})} \times
\frac{BR(B_{s} \rightarrow J/\psi \phi)}{BR(B_{u} \rightarrow J/\psi K^{+})}$
& $0.45 \pm 0.11 {\rm (stat)} \pm 0.07 {\rm (syst)}$ \\ \hline
$\frac{f(b \rightarrow B_{s})}{f(b \rightarrow B_{d})} \times
\frac{BR(B_{s} \rightarrow J/\psi \phi)}{BR(B_{d} \rightarrow
J/\psi K^{*}(892)^{0})}$ &
$ 0.27 \pm 0.07 {\rm (stat)} \pm 0.05 {\rm (syst)}$ \\
\hline \hline
$ \frac{BR(B_{d} \rightarrow J/\psi K^{0})}{BR(B_{u}
\rightarrow J/\psi K^{+})}$ &
$ 1.14 \pm 0.23 {\rm (stat)} \pm 0.08 {\rm (syst)} \pm ?
(b {\rm \: fraction}) $ \\ \hline
$\frac{BR(B_{d} \rightarrow J/\psi K^{*}(892)^{0})}{BR(B_{u} \rightarrow
J/\psi K^{+})}$ &
$1.68 \pm 0.34 {\rm (stat)} \pm 0.22 {\rm (syst)} \pm ?
(b {\rm \: fraction})$ \\ \hline
$\frac{BR(B_{s} \rightarrow J/\psi \phi)}{BR(B_{u} \rightarrow J/\psi K^{+})}$
& $1.13 \pm 0.28 {\rm (stat)} \pm 0.18 {\rm (syst)} \pm ?
(b {\rm \: fraction})$ \\ \hline
$\frac{BR(B_{s} \rightarrow J/\psi \phi)}{BR(B_{d} \rightarrow
J/\psi K^{*}(892)^{0})}$ &
$0.68 \pm 0.18 {\rm (stat)} \pm 0.13 {\rm (syst)} \pm ?
(b {\rm \: fraction})$ \\ \hline
$ \frac{BR(B_{u} \rightarrow J/\psi \pi^+)}{BR(B_{u}
\rightarrow J/\psi K^{+})}$ &
$0.049^{+0.019}_{-0.017} {\rm (stat)} \pm 0.011 {\rm (syst)}$ \\ \hline
\end{tabular}
\end{center}
\label{brratio}
\end{table}

\section{Rare Decays}

Rare $B$ decays provide a way to test the standard model against possible
effects due an anomalous magnetic moment of the $W^{\pm}$, a charged Higgs,
etc.  Such effects can contribute to the rate of
$B \rightarrow \mu^+ \mu^- K^{(*)}$ decays.  A nice way to measure the
rate of $B \rightarrow \mu^+ \mu^- K^{(*)}$ decays is to reference it
to the observed $B \rightarrow J/\psi K^{(*)}$ signals and use
theoretical input to give the expected relative rates and to
extrapolate from a limited $\mu^+ \mu^-$ mass region to the entire allowed
region.~\cite{carol}  CDF has done this in the Run~1a data
and obtained the following results (at 90\% confidence level):
\begin{displaymath}
\begin{array}[b]{rll}
BR(B^+ \rightarrow \mu^+ \mu^- K^+) < 3.5 \times 10^{-5} \\
BR(B^0 \rightarrow \mu^+ \mu^- K^*(892)^0) < 5.1 \times 10^{-5}
\end{array}
\end{displaymath}

These results are competitive with the similar limits from UA1 and CLEO.

The decay $B^0_{d,s} \rightarrow \mu^+ \mu^-$ also tests new particle
effects and is forbidden at the tree level.
Standard model predictions for the branching ratios are
$BR(B_d^0 \rightarrow \mu^+ \mu^-) = 8.0 \times 10^{-11}$ and
$BR(B_s^0 \rightarrow \mu^+ \mu^-) = 1.8 \times 10^{-9}$.~\cite{rare1}
Some extensions to the standard model predict that
$BR(B_s^0 \rightarrow \mu^+ \mu^-)$ can be as large as $10^{-8}$.~\cite{rare2}
Using
dimuon data with invariant mass between 4.9 and 5.8 GeV/c$^{2}$ in the
Run~1a sample, CDF finds
no $B^0_d \rightarrow \mu^+ \mu^-$ candidates in a mass window of
5.205 - 5.355 GeV/c$^{2}$ and 1 $B^0_s \rightarrow \mu^+ \mu^-$ candidate
between 5.300 - 5.450 GeV/c$^{2}$.  Normalizing to the
measured $B$ cross section, $\sigma(B^+) = 2.39 \pm 0.54 \: \mu b$ for
$p_T(B) > 6$ GeV/c and $\abs{y(B)} < 1$, and assuming
$\sigma(B^+) = \sigma(B_d^0) = 3 \sigma(B_s^0)$, we find (at 90\% confidence
level):
\begin{displaymath}
\begin{array}[b]{rll}
BR(B_d^0 \rightarrow \mu^+ \mu^-) < 1.6 \times 10^{-6} \\
BR(B_s^0 \rightarrow \mu^+ \mu^-) < 8.4 \times 10^{-6}
\end{array}
\end{displaymath}
These results are currently the best limits for these decay modes.~\cite{avery}

\section{Time-dependent $B_d$ Mixing}

The dimuon sample is also used to make the first time-dependent
mixing measurement at CDF.  Figure~10a illustrates the technique.
A vertex-finding algorithm is used to reconstruct a charm vertex in the
dimuon sample.  The combination of a well-measured charm vertex and one muon
can be used to reconstruct the decay length of the parent $B$ hadron,
where the charge of the muon gives the flavor of the $B$ hadron at decay.  The
charge of a second muon from another $B$ hadron indirectly gives the
flavor of the first $B$ hadron at production.  By measuring
the fraction of like sign muon events as a function of the reconstructed
pseudo-$c\tau$, the oscillation frequency for $B_d - \overline{B_d}$
mixing, $x_d$, can be determined.

After requiring the presence of two muons with $p_T(\mu) > 2$ GeV/c, a
reconstructed charm vertex, and $p_T^{rel}(\mu) > 1.3$ GeV/c, the final
data sample contains 1516 events with same sign muons and 2357 events with
opposite sign muons.  The $p_T^{rel}(\mu)$ cut significantly reduces the
sequential $b$ decay, $c-\overline{c}$, and fake lepton background
contributions.

A binned $\chi^2$ fit to the like sign muon fraction is used to extract $x_d$,
where the mixing from the $B_s$ component of the sample is assumed to be
maximal.  The $B$ lifetime, the relative fraction of $B_d$ and $B_s$ mesons
(from measured LEP values), and the background fraction are constrained in the
fit to within their Gaussian errors.  Most importantly, the time-dependent
contribution of
sequential $b$ decays to the like sign muon fraction is also included.
Detailed studies have shown that the kinematics
of the data sample agree well with the Monte Carlo sample used to
model the various background contributions.

Figure~10b shows the like sign muon fraction observed in the data compared with
the results for three different cases.  In the first case,
$x_d$ and $x_s$ are constrained to
be zero.  Here the expected contribution from sequential $B$ decay and other
backgrounds is shown.  In the second case, $x_d$ is set to zero again and the
contribution from maximal $B_s$ mixing is included.  The third case gives
the final fit results.
The CDF
preliminary measurement is $x_d = 0.64 \pm 0.18 {\rm (stat)} \pm 0.21
{\rm (syst)}$.  The modeling of the contribution from sequential $b$
decays is the dominant systematic uncertainty and this is expected to be
better understood with additional analysis.  This measurement has comparable
precision to previous results from LEP using a similar technique.  Additional
measurements of time-dependent mixing can be expected in the future using
other data samples.

\begin{figure}[thb]
\parbox[b]{3.0in}{\epsfxsize=3.0in\epsfbox{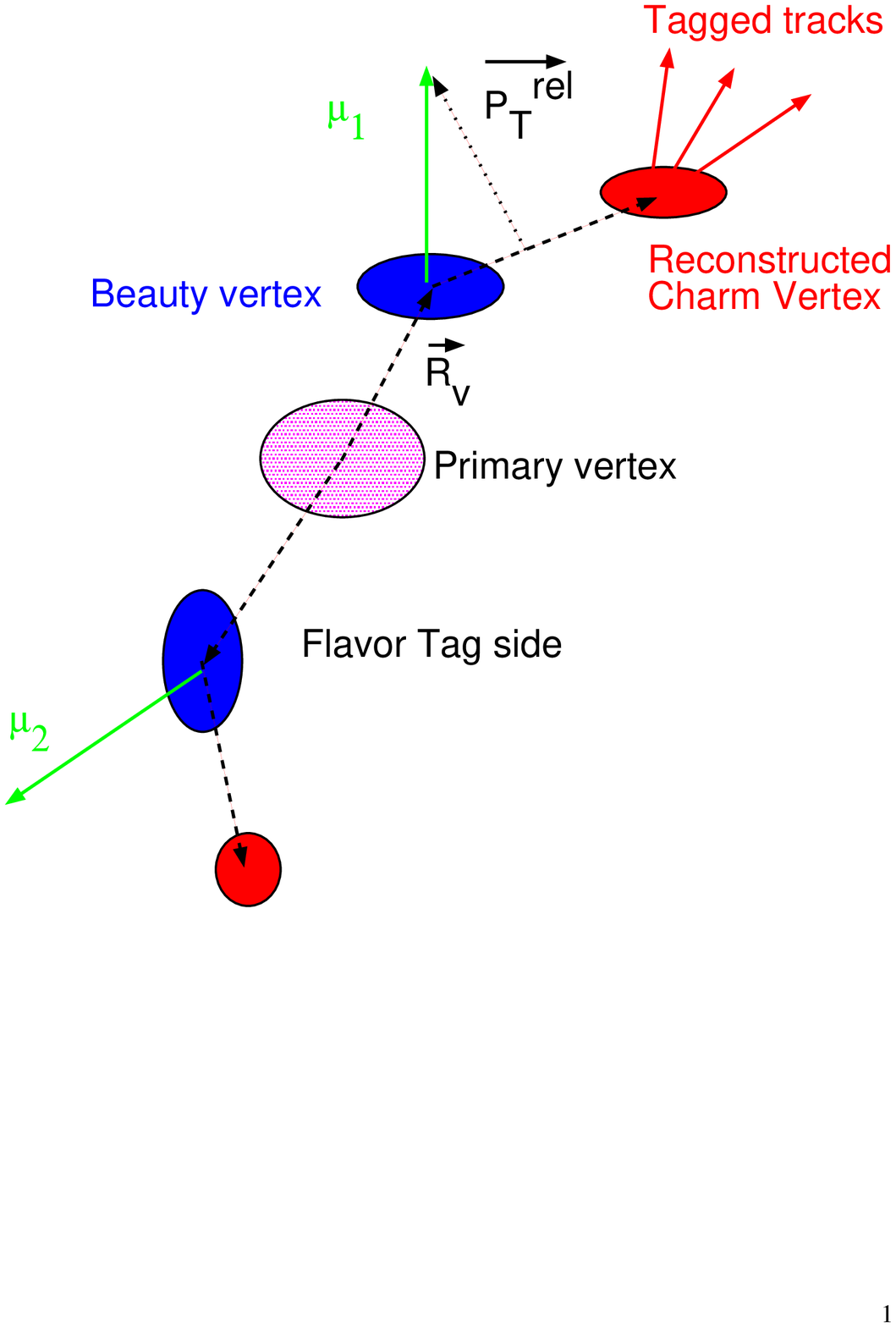}}%
\makebox[0in][r]{\raisebox{3.2in}{\em  a) \hspace{2.1in}}}
\hfill
\parbox[b]{3.0in}{\epsfxsize=3.0in\epsfbox{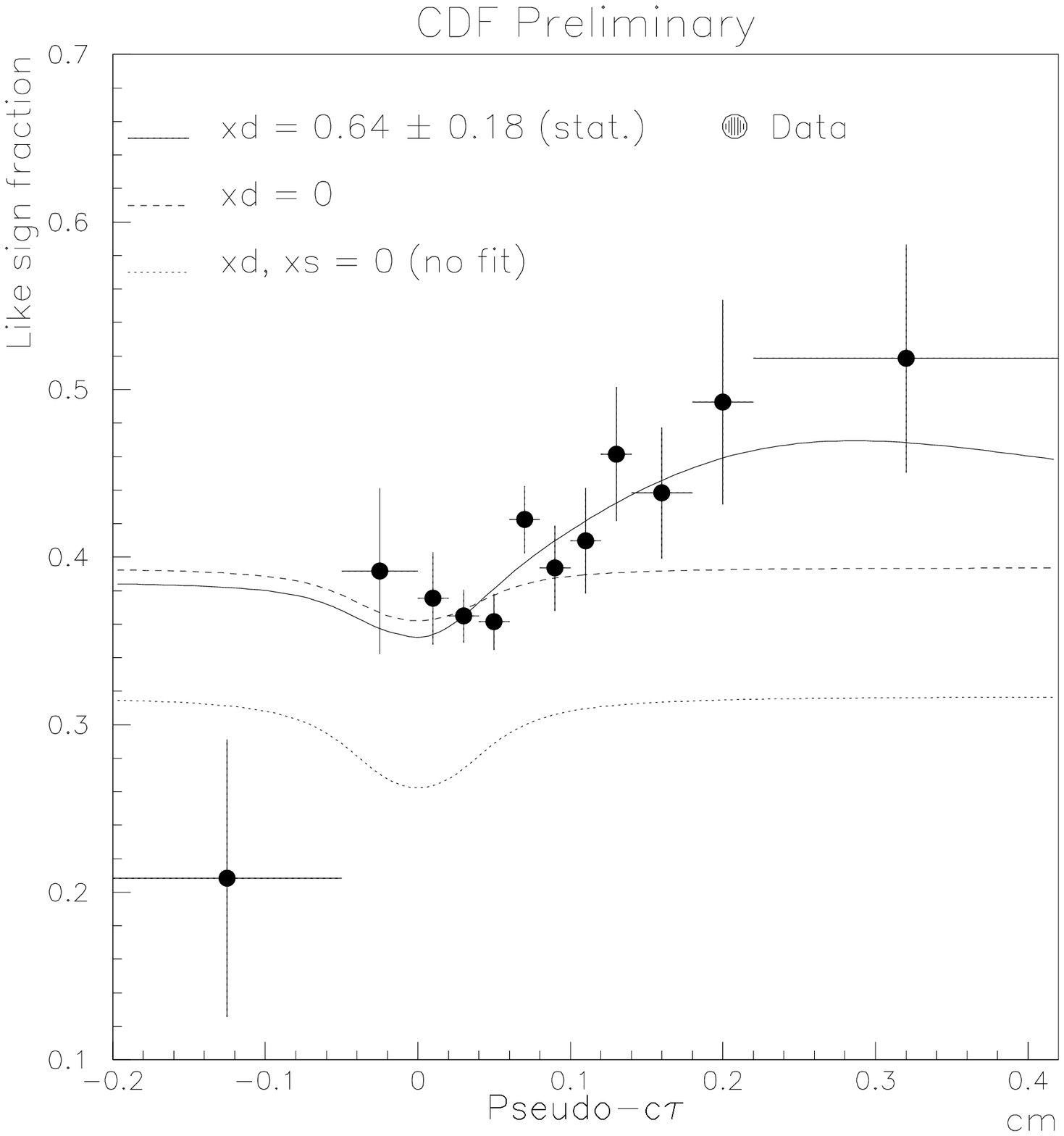}}%
\makebox[0in][r]{\raisebox{3.0in}{\em b) \hspace{2.1in}}}
\vspace{-0.3in}
\begin{center}
{\small Figure~10: a) Schematic picture of the technique used to measure
time-dependent $B_d$ mixing in the CDF dimuon sample. b) Like sign dimuon
fraction as a function of pseudo-$c\tau$.  The contributions from
sequential $b$ decays and background (dotted curve), the addition of
maximal $B_s$ mixing (dashed curve), and the observed $B_d$ mixing
(solid curve) are shown.}
\end{center}
\label{bfig12}
\end{figure}

\section{Conclusions}

The latest preliminary CDF results on $B$ decays have been presented.  The
addition of the Run~1b data will continue to improve all the CDF $B$ physics
results and allow new measurements so far not possible or expected.  This is
already seen in the $B$ lifetime results, the observation of the
Cabibbo-suppressed $B^+ \rightarrow J/\psi \pi^+$ decay, and in
time-dependent mixing measurements.
Results on $b$-baryons, $b$-tagging, $B_s$ mixing, and other exclusive $B$
decay modes will be coming soon to a conference and journal near you...
\bibliographystyle{unsrt}

\begin{thebibliography}{99.}

\bibitem{CDF} F. Abe et al. (CDF Collaboration),
{\it Nucl. Instrum. Methods Phys. Res., Sect. A}
{\bf 271}, 387 (1988), and references therein.

\bibitem{svx} D. Amidei et al.,
{\it Nucl. Instrum. Methods Phys. Res., Sect. A}
{\bf 350}, 73 (1994); P. Azzi et al., preprint FERMILAB-CONF-94/205-E.

\bibitem{thlife} V. Chernyak, preprint BudkerINP 95-18;
M. Shifman, preprint \\ TPI-MINN-94/32-T.

\bibitem{charge}  Throughout this paper, references to a specific charge state
imply the charge-conjugate state as well.

\bibitem{cdf_exlife} F. Abe et al. (CDF Collaboration),
{\it Phys. Rev. Lett.} {\bf 72}, 3456 (1994).

\bibitem{pdg}
L. Montanet et al. (Particle Data Group),
{\em Phys. Rev.} {\bf D50}, 1173 (1994).

\bibitem{blife_LEP} See talk by C. Stegmann at this workshop.

\bibitem{CLEO} CLEO Collaboration, {\it Phys. Rev.} {\bf D43}, 651 (1991).

\bibitem{morave} Moriond 95 lifetime averages from the LEP B Lifetimes Working
Group.

\bibitem{cdf_bslife} F. Abe et al. (CDF Collaboration), preprint
FERMILAB-Pub-94/420-E, accepted by {\it Phys. Rev. Lett.}

\bibitem{isi} I. Dunietz, preprint FERMILAB-PUB-94/361-T;
I. Bigi, preprint \\ UND-HEP-94-BIG06.

\bibitem{lewis} J. Lewis (For the CDF Collaboration), preprint
FERMILAB-CONF-94/128-E.

\bibitem{cleo1} CLEO Collaboration, preprint CLNS 94-1291.

\bibitem{carol} C. Anway-Wiese (For the CDF Collaboration), preprint
FERMILAB-Conf-94/210-E.

\bibitem{rare1} A. Ali, C. Greub, T. Mannel, Proceedings of the ECFA Workshop
on the Physics of the European $B$ Meson Factory (1993), 155.

\bibitem{rare2} J. L. Hewett et al., {\it Phys. Rev.} {\bf D39}, 250 (1989).

\bibitem{avery} See talk by P. Avery at this workshop.

\end{thebibliography}

\end{document}